# Simulating LTE and Wi-Fi Coexistence in Unlicensed Spectrum with ns-3

Lorenza Giupponi[a*], Thomas Henderson[b*], Biljana Bojovic[a], Marco Miozzo[a]

a. Centre Tecnològic de Telecomunicacions de Catalunya, Av. Carl Friedrich Gauss 7, Barcelona, Spain 08860
Email: {name.surname}@cttc.es

b. University of Washington, 185 Stevens Way NE, Seattle WA, 98195
Email: tomhend@u.washington.edu

* Equal contributors

*Abstract*— The use of unlicensed spectrum for future LTE systems raises concerns about its impact on co-located Wi-Fi networks. We report herein on extensions to a popular network simulator, known as ns-3, to model such coexistence. To our knowledge, this simulator is novel in two important ways. First, it is presently the only freely available simulator for coexistence studies, and as open source software, it offers new opportunities for reproducible research and collaborative model development. Second, we are not aware of other simulators with similar wireless models that allow full system studies to be conducted, such as performance evaluation of upper-layer protocols in a coexistence environment. We illustrate the value of a system simulator by describing a simulation campaign for the indoor scenario designed by 3GPP RAN1 using new ns-3 models for Release 13 LTE Listen Before Talk (LBT) techniques, intended for fair coexistence with Wi-Fi. LBT models have been widely simulated at the link and physical layer by many companies in 3GPP RAN1. We first show that we are able to tune the scenario to reproduce similar performance metrics to those reported in the literature. We then extend the simulations by adding TCP to the file transfers and explore the different behaviour observed. Our initial study reveals that there remain many open questions about how LBT should be implemented to achieve good coexistence properties, and that a broader scope of implementation issues, traffic models, and fuller system models should be studied further before drawing final conclusions on the LBT coexistence performance.

## I. INTRODUCTION

As wireless usage increases on expensive licensed bands, the mobile wireless industry is looking at offloads onto unlicensed bands. The evolution of LTE (Long Term Evolution) in 3GPP Release 13 and 14 is considering the so-called "Licensed-Assisted Access (LAA)" extensions to allow the licensed bands to be augmented by carriers located in unlicensed bands [1], through the principle of carrier aggregation.

The use of LTE in unlicensed spectrum generates multiple challenges [2], since LTE physical channels have been designed on the basis of uninterrupted and synchronous operation, while existing systems in the unlicensed bands operate in a decentralized, asynchronous manner employing protocols to promote fair usage of the spectrum. Therefore, a critical requirement for the design is that LTE coexists with other technologies like Wi-Fi in a "fair" and "friendly" basis, by extending its synchronous design. For that, a definition of "fairness" has been adopted by 3GPP [1]. Different regional regulatory requirements for transmission in unlicensed bands further complicate the design. In some markets, like Europe and Japan, a LBT (Listen-Before-Talk) function for clear channel assessment before accessing the channel is required, which implies changes to the LTE air interface. In other markets, such as North America, Korea and China, there are no such requirements. Techniques that enable coexistence with Wi-Fi can be realized for LTE Release 10/11/12 without changing the LTE air interface. For these specific regulatory requirements, the industrial consortium LTE-U (LTE in Unlicensed) Forum is specifying a proprietary solution, referred to as LTE-U [3]. To meet ETSI's LBT requirement, 3GPP has produced in Release 13 a new standardized version of LTE in unlicensed, LAA, for Supplemental Downlink (SDL) in unlicensed band. A Study Item has been recently finalized and has produced a Technical Report (TR) [1], where a summary of many simulation results is presented and discussed. Current efforts in Release 14 are focusing on defining also the uplink operation.

While different contributions in literature propose approaches to fairly share the spectrum [4]-[7], a key challenge to perform research on this topic is that despite the large body of simulation results in [1] and in the literature, the simulators are not publicly available. The obtained results are not easily reproducible, and system performance metrics are presented without much detail revealed about the underlying models and assumptions. To support LBT evaluation, Wi-Fi Alliance funded the development of simulation extensions to ns-3 [8], an open-source system simulator popular in research and academia. The objective of the project was to build an open simulator for evaluation of coexistence studies, to promote reproducibility, validation, and collaborative development. The requirement on the simulation study was to align with 3GPP RAN1 simulation scenarios, methodology and models.

Besides being openly available, ns-3 provides additional benefit to complement the existing body of work. Most existing coexistence simulators can be classified as link simulators with high fidelity models of the channel, physical layer and medium access control (MAC) layers, but high levels of abstraction at higher layers. In contrast, ns-3 is a full stack system simulator, with more abstraction than link simulators at the physical layer, but with higher fidelity models at higher layers. The ns-3 Wi-Fi models have been developed over time by several authors, usually by directly referencing the IEEE standards, starting with initial 802.11a models and later extending to many aspects of the 802.11b/g/p/e/n/ac standards. The ns-3 LTE models have been developed in close consultation with a small cell vendor and built around industrial small cell forum APIs; as a result, the models are product-oriented. Finally, ns-3 has some unique features at higher layers, including a real-time emulation mode, which allows code reuse on testbeds or real networks, and a capability to compile the source code of real applications and the Linux network kernel for direct use in the simulations. This capability dramatically reduces the gap between simulations and prototyping, allowing for code reuse in the area of LAA, where products are not yet in the market.

In this paper, we discuss an illustrative simulation campaign, we explore various sensitivities for the heavily parameterized LAA protocol and we highlight open research questions. We will discuss that, surprisingly, LAA access protocol parameter choices in general are not significantly impacting the coexistence performance. This is somewhat in line with the output of the 3GPP RAN1 evaluation. However, differently from the conclusion drawn in [1] and available literature, we show that other aspects can more strongly influence the coexistence behaviour, namely 1) the behavior of upper layer protocols such as TCP (Transport Control Protocol) and RLC (Radio Link Control), and 2) the channel occupancy and how traffic models influence it.

The outline of the paper is as follows. We introduce LAA and LBT novel concepts in Section II. We describe the models, scenarios and methodology that we reproduced following the specifics of 3GPP TR 36.889 [1] in Section III. We discuss an example of simulation campaign that can be carried out with the simulator in Section IV. We summarize our findings and suggestions for future work in section V, and the conclusions in Section VI.

## II. LAA AND LBT DESIGN

LAA is a system planned as a supplemental downlink in the 5 GHz unlicensed band, with the primary cell (PCell) always operating in a licensed band. 3GPP TR36.889 defines also as a key target the "fair" and "friendly" coexistence with Wi-Fi. The fairness is defined as the capability of an LAA network *not to impact Wi-Fi networks active on a carrier more than an additional Wi-Fi network operating on the same carrier, in terms of both throughput and latency*.

The LBT procedure is defined as a mechanism for a Clear Channel Assessment (CCA) check before using the channel, and is a regulatory requirement for the 5GHz unlicensed bands in Europe and Japan. The LBT uses, at a minimum, energy detection (ED) to determine if the channel is occupied. The availability of the channel cannot always be guaranteed, and certain regions impose limits on the maximum duration of a transmission burst.

Different options for LBT algorithms were evaluated by 3GPP, and the eventual algorithm selected was the one that bears most similarity to how Wi-Fi networks implement LBT (Category 4 LBT). Specifically, Wi-Fi implements a Distributed Coordination Function (DCF) or Enhanced Distributed Coordination Access (EDCA) algorithm that aims to resolve channel contention among competing nodes by implementing a random backoff with exponentially increasing maximum contention window and by imposing limits on the transmission opportunity (TXOP) before contention resolution occurs again. A state machine for the LBT CCA process is presented in [1]. Nodes wishing to transmit must observe a clear channel for an initial deferral period, after which if the equipment finds the channel to be clear, it may transmit immediately. On the other hand, if the medium is sensed to be already occupied, the transmission is deferred again and an extended CCA (ECCA) process is performed until the channel is deemed to be idle. In an ECCA check, the operating channel is observed for the duration of a random factor $N$ multiplied by the CCA slot duration. $N$ defines the number of clear idle slots that need to be observed before start of the transmission. The value of $N$ is randomly selected as $N \in [1,q]$ every time an extended CCA is required and the value is stored in a counter. The value of $q$ is the upper bound of the contention window, which varies according to an exponential backoff. The contention window size (CWS) is backed off upon collision detection, and reset upon absence of collision detection. In Wi-Fi, detection of a collision is performed using a control response (ACK) frame. In LTE, no such frame exists, so the collision detection is based on the hybrid ARQ (HARQ) feedback [9].

Figure 1 highlights the main additions we made to the ns-3 LTE and Wi-Fi models to enable coexistence. A new LBT Channel Access Manager was added to the LTE device, implementing the state machine described above, and hooked to the MAC and PHY layers of the LTE eNB model. We developed a new Wi-Fi PHY model compatible with the ns-3 multimodel spectrum framework, which allows Wi-Fi and LTE signals to coexist on the same channel. Wi-Fi and LTE reuse the interference managers and error models specific to their modules [8].

## III. METHODOLOGY, SCENARIO AND MODEL DESCRIPTION

We follow the 3GPP methodology for evaluation of fairness. Specifically, we consider two hypothetical operators, "operator A" and "operator B", using the same 20 MHz channel (Wi-Fi channel 36). We evaluate performance in two steps. In step 1, both operators deploy Wi-Fi technology. In step 2, operator A substitutes the Wi-Fi deployment with a LAA LBT network.

The operators deploy their networks according to two scenarios, indoor and outdoor, as designed and recommended for evaluation by 3GPP RAN1. We have implemented in the simulator both the recommended scenario, but due to space constraints, we focus herein on the 3GPP indoor scenario and configuration parameters, as discussed in [1]. Figure 2 provides an overview of the node layout. The two operators deploy four small cells in a building with no walls, and with dimensions as shown in Figure 2. The four base stations (BS) for each operator are equally spaced, but there is an offset on the x-axis. The UEs for both operators (20 each) are randomly distributed in the rectangular region, without redropping. Table 1 presents the details of the simulation scenario comparing the ns-3 implementation and the 3GPP model.

*A. Wi-Fi model*

We consider a 20 MHz 802.11n channel, with an Enhanced Distributed Coordination Access (EDCA) for best effort and voice traffic. Both AWGN and TGn fading channel model D are supported for the error model; results presented herein use the AWGN channel. The ED-based CCA for detection of other radio access technologies (RATs) implements a tunable ED threshold, which defaults to -62 dBm. Preamble detection (PD)-based CCA for Wi-Fi, allows for frame detection at the threshold of signal detection, around -88 dBm (i.e. more sensitive than the required -82 dBm threshold). This means that Wi-Fi will defer to weaker Wi-Fi signals sensed on the channel (down to roughly -88 dBm) compared to LTE signals (at -62 dBm threshold). Both beacons and RTS (Request to Send)/CTS (Clear to Send) are modeled. RTS/CTS functionality is not enabled in the results included in the paper, and the results are not sensitive to this choice. The current model is limited to 802.11n 2x2 MIMO (Multiple Input Multiple Output), supported by a MIMO abstraction model. Simulations described herein use rates up to Modulation and Coding Scheme (MCS) 15 with no short guard interval. An adaptive but idealized, feedback-based Wi-Fi rate control is used; rate control adjustments are made immediately upon feedback from the peer and not due to a probing algorithm such as Minstrel.

*B. LAA model*

LAA implements a LBT protocol. All LBT parameters were approved in 3GPP RAN Plenary meeting in December 2015 [9], after having been agreed in the context of 3GPP RAN1. Initial and extended CCA defer at 43 μs, and the LAA CCA slot time is 9 μs [9]. LAA ED threshold is tunable separately from Wi-Fi's threshold, and its value is set to -72 dBm. The maximum length of TxOP is configurable and it defaults to 8 msec. The update of the CWS is implemented following a HARQ feedback based approach, as agreed in [9][10]. The HARQ based rule declares a collision, and consequently updates the CWS, if Z=80% of feedbacks from the first subframe of the latest transmission burst are NACKs. Otherwise, the CWS is reset to the minimum value (i.e. 15). The upper bound of the contention window varies according to a Category 4 exponential backoff between {15, 31, 63}. Data transfer starts at the subframe boundary. We implement reservation signals to occupy the channel until the subframe boundary, to force other nodes to defer while LAA is not yet occupying the channel with data. The reservation signals count against the node's maximum allowed TxOP time. Discovery Reference Signal (DRS) signals are sent during the so called Discovery Signal Measurement Timing Configuration (DMTC) window (6 msec between subframe 0 and 5), with a tunable periodicity of 40/80/160 msec. The periodicity defaults in our tests to 80 msec. DRS transmission is subject to a priority LBT with a fixed defer period of 25 msec. If data is scheduled during the DTMC window, DRS is embedded with data, otherwise it is sent alone without data, and modeled as a transmission occupying 1 msec. The system information (MIB/SIB1) is channeled through the PCell. We support MIMO with both spatial and transmission diversity, and up to MCS 28. The simulator offers standard compliant [8][12] implementations of RLC-UM (unacknowledged mode), RLC-AM (acknowledged mode) and RLC-TM (transparent mode).

*C. Traffic Models*

The overall offered load is the same for both coexisting networks. TR36.889 [1] calls for several traffic models, as shown in Table 1. In ns-3, we have implemented the FTP (File Transfer Protocol) Model 1 for this project, and evaluated it on a downlink only scenario, as one of the recommended options in [1], according to different arrival rates λ. We implement it on top of IP and either UDP or TCP. This model simulates file transfers arriving according to a Poisson process with arrival rate lambda. The recommended range for λ is between from 0.5 to 2.5. The file size is 0.5 Mbytes. We also implemented a voice model corresponding to [1], that when enabled, substitutes the file transfer for two downlink nodes on operator B's network with a voice flow that is measured for latency and outage. In addition, we support a constant bit rate UDP traffic model option, similar to voice models, with varying bit rates up to saturation.

*D. Performance metrics*

The main performance metrics described in TR 36.889 [1] are 'user perceived throughput' and 'latency'. In ns-3, we are calculating these by using the built-in FlowMonitor tool that tracks per-flow statistics at the IP layer including throughput and latency, and we then post-process these flow results to obtain CDFs. We also track latency separately between voice and non-voice flows. Due to space considerations, only throughput metrics are shown herein. The simulator also allows to log and classify all physical layer transmissions (duration and power), backoff values drawn, evolution of the contention window, HARQ feedback logs, Wi-Fi retransmissions, and TCP retransmission events.

IV. SIMULATION RESULTS AND DISCUSSION

After interactions with different vendors belonging to Wi-Fi Alliance and 3GPP, we learned that the simulators used in 3GPP RAN1 for evaluations do not consider full protocol stack models, but are mainly focused on PHY-MAC models. The FTP application is therefore modeled as raw data sent over the MAC. To model this we configured the FTP application to operate over UDP and RLC-UM, so as to avoid retransmissions at both link and transport layers. However, in the current Internet, file transfers are typically run over TCP. TCP commonly runs over a reliable link, supported by RLC-AM in LTE networks. We therefore further tested the same FTP application over TCP and RLC-AM.

We have simulated different loads by tuning the traffic intensity λ values, mainly within the range specified by 3GPP (between 0.5 and 2.5). We scaled the simulation duration depending on the offered traffic load to try to obtain similar numbers of flows despite different traffic intensity, and also configure a brief warmup phase at the start. In the figures that follow, we chose to illustrate results for λ=2.5, with simulation duration of 384 seconds, because lower traffic intensities do not occupy the link enough to show interesting performance differences. Finally, we note that all of the results contained in this section can be reproduced [11].

### A. FTP Model 1 over UDP (and RLC UM)

The throughput impact on the non-replaced Wi-Fi network is shown in Figure 3b. In Step 1, the Wi-Fi networks occupy the channel during approximately 10% (combined) of time. On the other hand, in Step 2, when operator A is replaced by a LAA system, Wi-Fi occupies the channel for 5.7% of time, and the LAA system occupies the channel for a comparable amount (5.2%). We observe that although there is a small amount of throughput impact (Figure 3b) as compared with the replaced Wi-Fi network (Figure 3a), the user impact is likely negligible. This finding is consistent with the body of work reported in [1].

The coexistence performance is affected by different aspects and LAA parameters, and we have studied the impact of many of them. We discuss some examples in the following. The simulator offers opportunities for further interesting studies along these lines.

*Impact of control signals.* Besides the reservation signals, additional control data is sent in the unlicensed band. The PDCCH (Physical Downlink Control Channel), carrying DCI (Downlink Control Indicator), occupies between 1 and 3 symbols per subframe. We consider 2 symbols in our simulations. In addition, Radio Resource Management (RRM) features require the DRS to be sent every 40, 80 or 160 msec. In case the DRS signals have to be sent without data, they may require more channel airtime than the corresponding Wi-Fi beacons, each of which occupies 0.176 msec airtime, with periodicity of 100msec. We tested 40, 80 and 160 msec DRS intervals and we found that LAA would occupy between 4.4%, 5.2% and 6.8% of the airtime, in the three cases, in comparison with 4.7% of Wi-Fi network in Step 1. Coexistence performance is represented in Figure 4a. High periodicity of DRS signals would be beneficial for RRM measurements, but may impact the coexistence performance by requiring more user data transmissions to defer.

*Impact of LBT protocol.* We have tested the impact of the parameters associated with the LBT access protocol and with the backoff algorithm. The constants specified in these algorithms were the subject of many parametric sensitivity studies in 3GPP. In particular, we tested the sensitivity to 1) the Z parameter associated to the HARQ based rule to update the contention window size, 2) the maximum contention window size, and 3) the maximum length of the LBT TxOP. Data for these results are not shown here due to space constraints, but information can be found in [11]. In general, we observed that performance is not much affected by these parameters, which is a result consistent with the output of the simulation studies contained in [1]. On the other hand, it may also mean that key aspects of the LBT design, such as the technique to update the CWS are not properly defined.

In particular, we are concerned about the HARQ rule for collision detection in the LAA system. While Wi-Fi feedback after a transmission is received in a matter of microseconds, the HARQ feedback is received with much more delay, e.g. after 7 msec. This delay may cause the feedback to become outdated. In addition, differently from Wi-Fi, where the sender receives only one feedback per TxOP, the eNB receives multiple HARQ feedbacks from each UE served during the TxOP, and this information has to be combined based on the Z threshold rule [10]. Finally, only feedback associated to the first subframe of the TxOP is considered, based on [9], which means that a collision happening in a different subframe will be ignored. Overall we observe that increased collision probability leads Wi-Fi to face longer backoff times than LAA.

*Impact of hidden terminals.* We have evaluated the coexistence performance also in a modified indoor scenario, where the BSs are located at the corners of the simulation scenario. This increases the average distance between users and BSs, and also the number of hidden nodes in the scenario. We recall that LAA nodes are energy detected by WiFi at -62 dBm, while WiFi nodes are energy detected by LAA at -72 dBm, and WiFi nodes preamble detect each other at -88 dBm. These asymmetric detection levels make that LAA is more affected than Wi-Fi by the increased number of hidden nodes in the scenario, as it is shown in Figure 4b. It can be observed how this performance loss can be reduced if the Wi-Fi detection threshold is lowered to e.g. -82 dBm. An option considered in real products is CTS2self [15] support for LAA, which would enable Wi-Fi to preamble detect also LAA at -82 dBm or lower.

### B. FTP Model 1 over TCP

When considering TCP, performance is affected by the same aspects we observed with UDP, but we also found a more substantial impact on Wi-Fi. Coexistence performance in Steps 1 and 2, in terms of per flow throughput, are shown in Figure 5. Due to typical protocol stack delays [14], each round trip time (RTT) in LAA takes about 15 msec or longer if buffers grow, which bounds the throughput per flow to approximately 20 Mbps.

We configured a segment size of 1440 bytes and an initial congestion window of ten segments, with TCP NewReno congestion control. In this case, when in Step 1, the combined Wi-Fi networks are occupying the channel during 11% of the time, in Step 2 the LAA system occupies the channel during 13% compared with the non-replaced Wi-Fi network's channel occupancy of 6% of time. This increased airtime occupancy generates more contention and more collisions, which increase from 1.4% in step 1, to 8% in step 2. There are different reasons for this increase in channel occupancy.

*Impact of LTE resource allocation granularity.* Unlike Wi-Fi, LAA channels are structured around 1 millisecond subframes. Small amounts of data traffic, such as those transmitted when the congestion window is small, may occupy the full millisecond transmission, compared with tens of microseconds for the equivalent Wi-Fi transmissions. These inefficiencies in the subframe occupancy should be tackled at scheduler level.

*Impact of TCP congestion window*. The coexistence of two windows, TCP and RLC, generates a flow control effect that can alter the data arrival pattern as compared with the bursty arrival rate observed when FTP is run over UDP. This means that TxOPs are not completely packed, which is expensive in terms of channel occupancy. The initial congestion window value also influences the number of initial round trips in which TCP cannot send enough data to fill much of the subframes.

*Impact of upper layer protocols overhead*. RLC-AM introduces additional overhead in terms of STATUS-PDU and retransmissions, if they are not scheduled in PCell.

## V. SUMMARY OF FINDINGS AND SUGGESTIONS FOR FUTURE WORKS

In the previous section we have only touched upon the potential of the simulator and some of the simulation results that we have obtained, but we wish to stress the following results and general observations:

- Coexistence performance is highly sensitive to factors that affect the channel occupancy (e.g. control signals), even more than to the parameter choices in the LBT CCA and backoff algorithms.
- Channel occupancy, and consequently coexistence, is not only affected by the behavior of the PHY-MAC layers, and of the LAA access in particular, which have been evaluated in 3GPP RAN1 and literature, but also by other aspects, related with upper layer protocols, such as TCP and RLC. However, no other previous study has included evaluations of TCP performance, to our knowledge.
- A bursty traffic pattern, such as the FTP run over a UDP or raw transport, may be a best-case scenario for coexistence in LAA scenarios, because inefficiencies of LAA in accessing the channel, due to the resource allocation granularity of 1 msec, can be amortized when transmissions are bursty. Other less bursty traffic models, or other transport protocols, e.g. TCP, may cause LAA to occupy the channel more frequently and inefficiently and impact the coexistence with other technologies.
- Concerns have been highlighted with respect to LBT basic design aspects such as the HARQ-based CWS update. These design choices may be responsible for the fact that no significant impact in coexistence is observed when modifying these LBT parameters.
- Either CTS2self, which allows Wi-Fi to preamble detect LAA, or support for lower Wi-Fi energy detection thresholds, seems to be a fundamental functionality to be supported by LAA/Wi-Fi to allow coexistence with Wi-Fi, and to protect the LAA performance in the presence of hidden nodes.

For future research, we recommend the design of smart scheduling approaches capable of solving the inefficiency and granularity issues in LAA resource allocation, which are highlighted in cases of applications run over TCP, or in the case of constant bit rate applications, such as voice. In addition, we recommend investigating the effectiveness of the HARQ based collision detection approach.

Finally, with respect to the development roadmap of the simulator, it includes the extension to carrier aggregation functionality, and the comparison to LTE-U approaches. In addition, it may include future evolutions of LAA related with uplink transmissions, i.e. Release 14 eLAA, as well as the MuLTEFire technology to develop LTE entirely in unlicensed band, without an anchor in the licensed band [13].

## VI. CONCLUSION

In this paper we have described extensions to the discrete event network simulator ns-3 to enable evaluation of LAA-Wi-Fi coexistence performance on a full protocol stack model. We believe that this platform is unique, complements the large body of LAA simulation results with link simulators, and allows for new insights in coexistence evaluation, with respect to previous works presented in literature and especially in 3GPP TR36.889. Our models and scripts to reproduce this work are freely available to other researchers. We have run a simulation compaign aligned with the methodology proposed by 3GPP RAN1, we have studied parameter sensitivites and we have shown that when considering full protocol stack and high fidelity models for both LAA and Wi-Fi, many behaviors are observed, which were not highlighted in the existing body of work. This paves the way to new research topics that will more fully answer under what conditions LAA is capable of fairly coexisting with Wi-Fi.

## VII. ACKNOWLEDGEMENTS

This work has been supported by Wi-Fi Alliance. The authors would like to thank Nicola Baldo for developing scenario and propagation models during an earlier phase of this project, and Benjamin Cizdziel and Hossein Safavi for developing Wi-Fi error rate models. The authors also acknowledge helpful discussions with Sumit Roy and Farah Nadeem, as well as modelling discussions with several vendors.

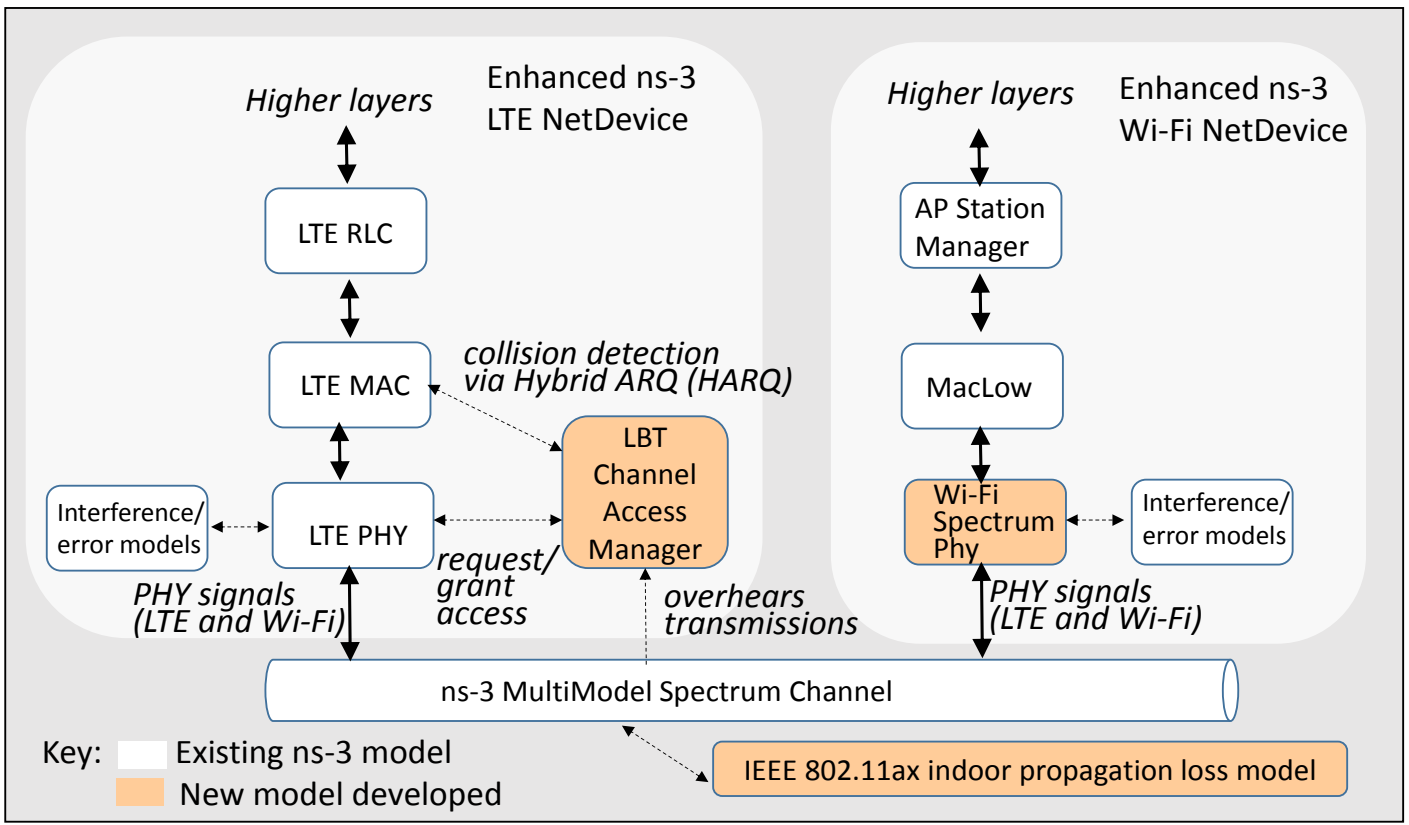

Figure 1 Block diagram of coexistence simulator

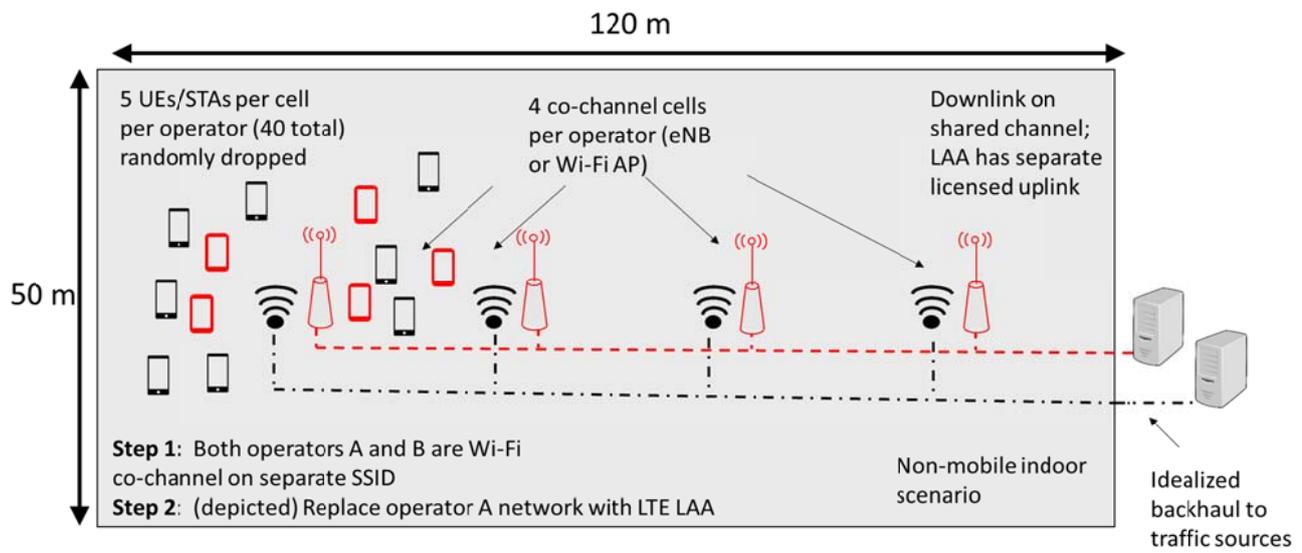

**Figure 2 Indoor scenario**

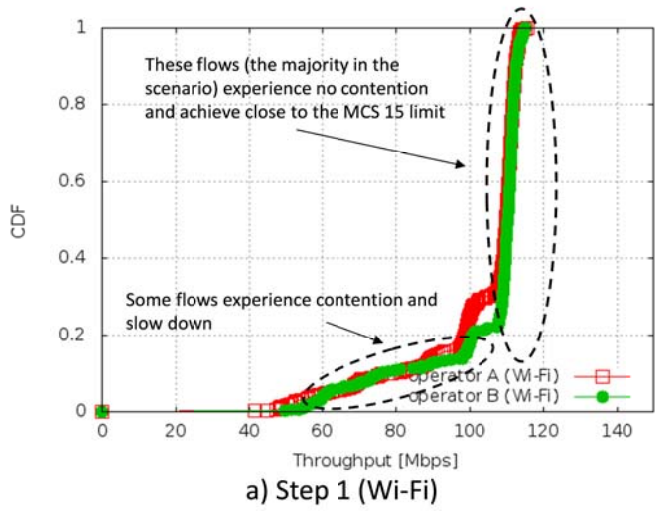 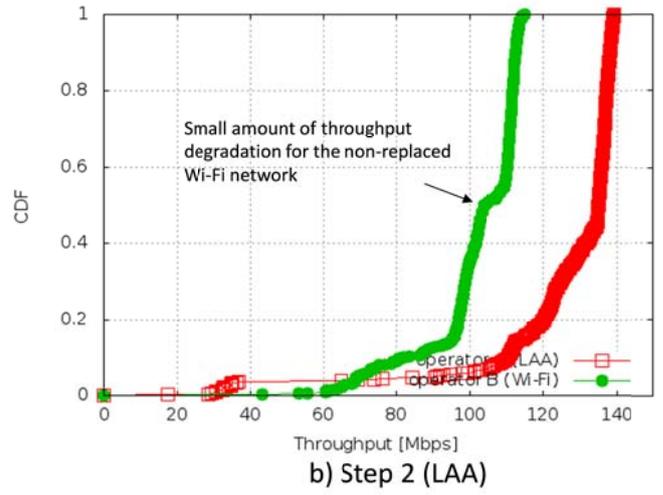

Figure 3 Throughput performance: FTP (arrival rate λ=2.5) over UDP and RLC-UM

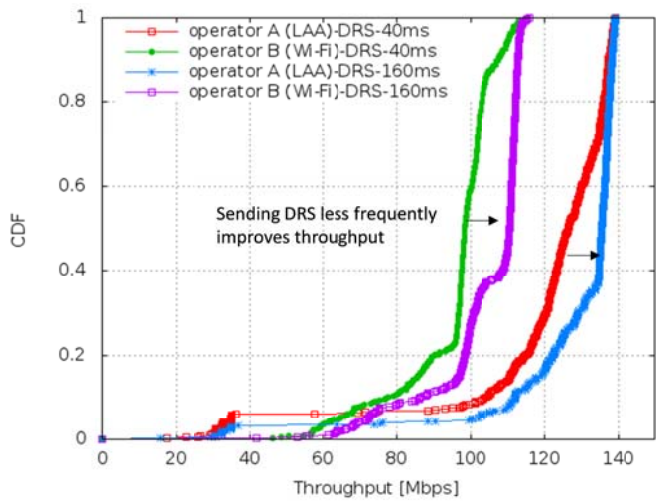 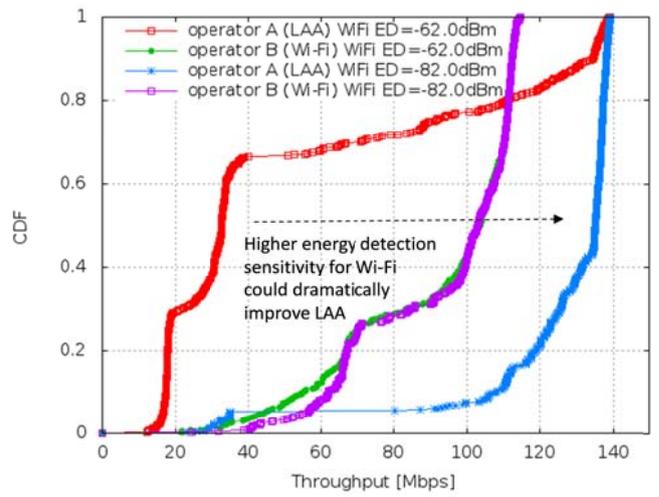

**Figure 4 Throughput performance sensitivity to sample parameters**

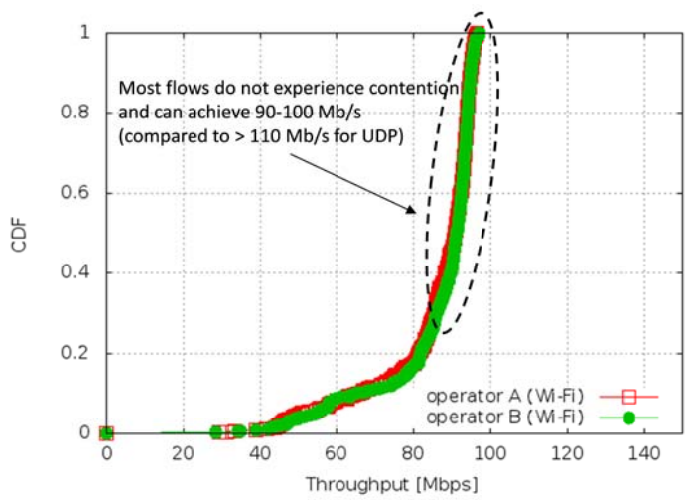 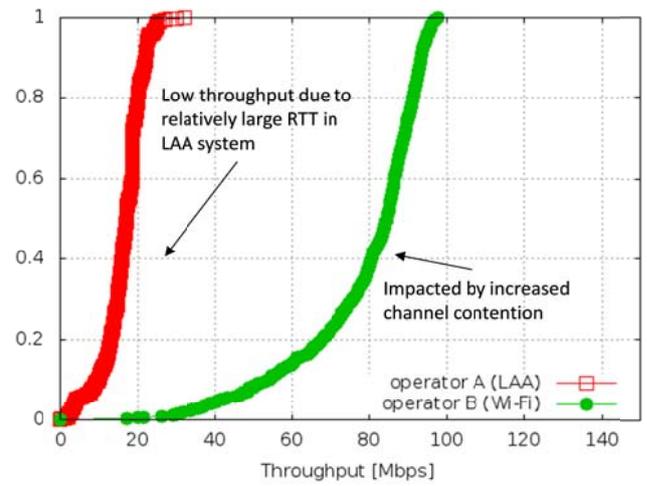

**Figure 5 Throughput performance: FTP (arrival rate λ=2.5) over TCP and RLC-AM**

Table 1 Scenario comparison between 3GPP TR 36.889 and ns-3 implementation

| Unlicensed channel model | 3GPP TR 36.889 | ns-3 implementation |
|---|---|---|
| Network Layout | Indoor scenario | Indoor scenario |
| System bandwidth | 20 MHz | 20 MHz |
| Carrier frequency | 5 GHz | 5 GHz (channel 36, tunable) |
| Number of carriers | 1, 4 (to be shared between two operators) 1 for evaluations with DL+UL Wi-Fi coexisting with DL-only LAA | 1 for evaluations with DL+UL Wi-Fi coexisting with DL-only LAA |
| Total BS transmission power | 18/24 dBm | 18/24 dBm Simulations herein consider 18 dBm |
| Total User equipment (UE) transmission power | 18 dBm for unlicensed spectrum | 18 dBm |
| Distance dependent path loss, shadowing and fading | ITU InH | IEEE 802.11ax indoor model |
| Antenna pattern | 2D Omni-directional | 2D Omni-directional |
| Antenna height | 6 m | 6 m (LAA, not modelled for Wi-Fi) |
| UE antenna height | 1.5 m | 1.5 m (LAA, not modelled for Wi-Fi) |
| Antenna gain | 5 dBi | 5 dBi |
| UE antenna gain | 0 dBi | 0 dBi |
| Number of UEs | 10 UEs per unlicensed band carrier per operator for DL-only 10 UEs per unlicensed band carrier per operator for DL-only for four unlicensed carriers. 20 UEs per unlicensed band carrier per operator for DL+UL for single unlicensed carrier. 20 UEs per unlicensed band carrier per operator for DL+UL Wi-Fi coexisting with DL-only LAA | Supports all the configurations in TR 36.889. Simulations herein consider the case of 20 UEs per unlicensed band carrier per operator for DL LAA coexistence evaluations for single unlicensed carrier. |
| UE Dropping | Randomly dropped and within small cell coverage. | Randomly dropped and within small cell coverage. |
| Traffic Model | FTP Model 1 and 3 based on TR 36.814 FTP model file size: 0.5 Mbytes. Optional: VoIP model based on TR36.889 | FTP Model 1 as in TR36.814. FTP model file size: 0.5 Mbytes Voice model: DL only |
| UE noise figure | 9 dB | 9 dB |
| Cell selection | RSRP for LAA UEs and RSS for Wi-Fi STAs | RSRP for LAA UEs and RSS for Wi-Fi STAs |
| Network synchronization | Small cells are synchronized, different operators are not synchronized. | Small cells are synchronized, different operators are not synchronized. |